 \documentclass{ametsocV6.1}

\authors{Johnson Dhanasekaran,\aff{a}
Donald. L. Koch,\aff{b} \correspondingauthor{Donald. L. Koch, dlk15@cornell.edu} 
}
\affiliation{\aff{a}{Sibley School of Mechanical and Aerospace Engineering, Cornell University, Ithaca, New York 14853, USA}\\
\aff{b}{{Smith School of Chemical and Biomolecular Engineering,Cornell University, Ithaca, New York 14853, USA}}\\
}

\title{The effect of turbulence, gravity, and non-continuum hydrodynamic interactions on the drop size distribution in clouds}

\abstract{The evolution of micron-sized droplets in clouds is studied with focus on the 'size-gap' regime of 15-40 $\mu m$ radius, where condensation and differential sedimentation are least effective in promoting growth. This bottleneck leads to inaccurate growth models and turbulence can potentially rectify disagreement with in-situ cloud measurements. The role of turbulent collisions, mixing of droplets, and water vapour fluctuations in crossing the 'size-gap' has been analysed in detail. Collisions driven by the coupled effects of turbulent shear and differential sedimentation are shown to grow drizzle sized droplets. Growth is also promoted by turbulence-induced water vapour fluctuations, which maintain polydispersity during the initial condensation driven growth and facilitate subsequent growth by differential sedimentation driven coalescence. The collision rate of droplets is strongly influenced by non-continuum hydrodynamics and so the size evolution beyond the condensation regime is found to be very sensitive to the mean free path of air. Turbulence-induced inertial clustering leads to a moderate enhancement in the growth rate but the intermittency of the turbulent shear rate does not change the coalescence rate significantly. The coupled influence of all these phenomena is evaluated by evolving a large number of droplets within an adiabatically rising parcel of air using a Monte Carlo scheme that captures turbulent intermittency and mixing.}

\begin{document}

\maketitle

%
%
%
\statement
This study is directed toward improving descriptions of the microphysical determinants of the time for rain formation in clouds. Existing models predict significantly longer times than the tens of minutes observed in warm clouds. There is a growing body of evidence that turbulence plays a key role in resolving this discrepancy. We incorporate accurate turbulent collision dynamics and assess the interplay of the various underlying physical factors facilitating growth to rain-sized droplets. Our study, in addition to providing important insight into cloud microphysics, will pave the path to the next generation of large-scale rain cloud evolution studies.

%
%
%

%

\section{Introduction}\label{sec:introduction}

The size distribution of water droplets in clouds affects radiative heat exchange while its evolution controls precipitation and so it is an important component in the study of our atmosphere \citep{peng2002cloud,grabowski2013growth}. In clouds small droplets, typically with radii smaller than 15 $\mu$m radius, grow by condensation while large ones, typically with radii greater than 40 $\mu$m, collide through differential sedimentation to coalesce. However, the growth rate in the 'size-gap', of 15-40 $\mu$m radii, is not fully understood and it leads to significant errors in predictions of the time to grow drizzle-sized droplets and increase the width of the size-distribution\citep{chaumat2001droplet,grabowski2013growth}. To reconcile models to in-situ cloud measurements various mechanisms have been proposed and we will study in detail the role of turbulence in inducing polydispersity and collisions, the effect of non-continuum hydrodynamic interactions on collision  rate, and the collision frequency enhancement due to inertial clustering. To obtain qualitative and quantitative estimates of their importance we evaluate the droplet distribution evolution.

The growth of small droplets is controlled by condensation. We simulate it within an adiabatic air parcel whose upward motion changes thermodynamic quantities, such as temperature and water vapour pressure, leading to diffusion of water vapour onto the droplets. Acting in isolation this effect would create a monodisperse distribution,  making the relative velocity of drops due to gravity vanish. Additionally, condensation cannot grow very large droplets and so droplets cannot generate enough relative velocity by their own weight to drive collision. This leads to the 'size-gap' and we test the role of the aforementioned mechanisms in breaking through it.

For collision-induced growth of droplets through the 'size-gap', turbulence is expected to play a role \citep{kostinski2005fluctuations,chandrakar2024turbulence} along with its coupling with differential sedimentation. Thus we use the collision rates calculated by \cite{dhanasekaran2021collision_a} that properly account for the trajectories of hydrodynamically interacting drops resulting from the simultaneous driving forces. \cite{li2018effect} perform direct numerical simulations (DNS) with particles settling in turbulence but did not incorporate hydrodynamic interactions which have a strong influence on the relative motion of droplets as they approach each other. Studies by \cite{xue2008growth} and \cite{grabowski2009diffusional} include hydrodynamic interactions but without properly coupling turbulence and gravity. In addition these studies do not include the breakdown of continuum as the droplets approach each other which is expected to be the predominant mechanism allowing drops to overcome the lubrication resistance to drop-drop contact.  For a large range of droplet sizes these non-continuum effects predominate over van der Waals forces and surface mobility as mechanisms facilitating coalescence \citep{sundararajakumar1996non,dhanasekaran2021collision_b}. Finally high Reynolds numbers based on the Taylor microscale (Re$_{\lambda}$), of $\textit{O}(10^4)$ typical in clouds, have not been reached in the DNS studies by \cite{li2018effect} and this can lead to errors of up to 20$\%$ in the observed collision rate. All of these issues have been properly resolved in the collision rate calculated by \cite{dhanasekaran2021collision_a}, which will be used in our evolution study.

Droplets possess inertia causing them to lag the background flow. For conditions typical in clouds droplet response times vary from very small to comparable with the characteristic fluid time, corresponding to the Kolmogorov time scale for sub-Kolmogorov droplets\citep{ayala2008effectsa}. In these conditions droplet inertia enhances the pair probability density or the local concentration of neighboring droplets, thereby increasing the collision rate. Inertia causes the trajectories of drops with larger response times to detach from the local flow allowing them to collide with other drops having a different turbulence history. This process, referred to as the 'sling-shot effect' leads to 'caustics' \citep{wilkinson2006caustic,falkovich2007sling}.  However no comprehensive collision rate predictions are available in this regime which affects a small fraction of droplets and so it will be not be considered in our study. Inertial clustering depends on the bidispersity of the pair of interacting drops and the coupling of gravity with turbulence. We use the analytical expression from \cite{dhanasekaran2022model} that spans a large parameter space, pertinent to the droplets of interest, and reproduces reported DNS results as well as important physics. The enhanced local concentration, which increases the likelihood of collision events, will be incorporated into our droplet evolution model.

The turbulent mixing of water vapour in the presence of gradients in vapour partial pressure leads to  fluctuations of the vapour pressure. These fluctuations have Gaussian statistics and can play an important role in droplet evolution by generating polydispersity\citep{kulmala1997effect,chandrakar2016aerosol,li2019cloud}. Broadening of the drop size distribution is also expected via mixing of droplets\citep{lasher2005broadening,siebesma2003large}. We model both of these mechanisms using packets, with any given packet representing a collection of droplets with a unique history of turbulent velocity gradients and water vapour pressures. The water vapour fluctuation in any particular packet is approximated as a stochastic mean-reverting process with a standard deviation typically about 1$\%$ of the mean\citep{kulmala1997effect,chandrakar2016aerosol} that can however be attenuated by condensation. Mixing of droplets is modelled by moving them between packets. Since mixing is governed by motion of the large scale eddies both the rate of reversion of the turbulent water vapour fluctuations and the rate of droplet mixing are chosen to be the inverse of the integral time scale of the turbulence.

There are a large number of droplets, about $10^8$ per $m^3$, in a typical cloud\citep{li2017eulerian}. It is not numerically  feasible to track all of them while still accounting for all the mechanisms in play. Thus, we use a Monte Carlo scheme with one Monte Carlo droplet representing many real droplets to retain the discrete nature of the drop size distribution. This discrete distribution, relative to a continuous one, better captures the stochastic growth driven by the various modes of turbulence and is complemented by the use of packets to capture different realisation of turbulence. The collection of all the packets represents the adiabatic air parcel rising from the ground.

The model developed here focuses on growth of water droplets in turbulent clouds. However, it can be easily extended to other cases such as industrial aggregators. For industrial aggregators, \cite{buesser2012design} discuss the critical role of particle collision in growing carbon black, pigments, and other commercially valuable products when starting from an initial phase controlled by diffusional growth.  The aerosol in these aggregators typically experiences a mean acceleration that acts as an effective body force.  Thus, this system mimics the droplets in clouds with the effective body force on the particles taking on the role of gravity. The high temperatures of the operation could potentially lead to large enough mean free paths for the turbulent gas that collisions of the sub-micron particles are governed by non-continuum effects.

In this study we first develop the Monte Carlo method and all pertinent formulations for the cloud packet model in \S\ref{sec:formulation}. We then utilise it to obtain the evolution of the drop size distribution. Important results and insights obtained from this calculation are presented in \S\ref{sec:results}. Finally in \S\ref{sec:conclusion} we summarize and highlight some of the important insights of our study.

\section{Formulation}\label{sec:formulation}

Water droplets in clouds typically initially grow through condensation onto sub-micron aerosol particles. It is possible that a few very large, about tens of micron radii, nucleation sites could impact the evolution in the 'size-gap' and this has been studied by \cite{feingold1999impact} and \cite{lasher2001early}. However, for the vast majority of nuclei, condensation along with  turbulent water vapour fluctuations will control the growth rate and the shape of the distribution to sizes in the lower end of the 'size-gap' \citep{kulmala1997effect,chandrakar2016aerosol,li2019cloud}. Hence we include condensation in our model to obtain an accurate description of the drop size evolution and do not focus on details of the initial cloud condensation nuclei.

To model condensation we simulate an adiabatic parcel of air rising from the ground starting with initial  relative humidity of 100$\%$ at a typical updraft velocity of $1\, m/s$ \citep{warner1970microstructure}. As it rises, due to dropping temperature and pressure, the water vapour carrying capacity decreases and the vapour diffuses to and condenses on droplets. For a droplet of radius $a$ the growth rate is given as,
\begin{eqnarray}
\frac{da}{dt}=\frac{D}{\rho \, a} \left[\rho_{\infty}-\rho_{a}(T_a) \right]
\label{eq:condensation}
\end{eqnarray}
Here, $D$ is the diffusivity of water vapour, $\rho$ is the density of water , $\rho_{a}(T_a)$ is the water vapour density in local equilibrium with the surface of the droplet at its temperature $T_a$, and $\rho_{\infty}$ is water vapour density of air in the vicinity of the droplet. We can calculate $\rho_a$ using the Clausius–Clapeyron relation. The temperature of the adiabatic parcel $T_{\infty}$ is tracked by accounting for latent heat lost to condensation and the adiabatic cooling due to expansion as the packet rises. On the other hand $T_a$ is determined by an energy balance on the drop given by,
\begin{eqnarray}
\frac{dQ}{dt}=4\pi a [T_a - T_{\infty}] K_c 
\label{eq:heat_conduction}
\end{eqnarray}
Here $Q$ is the energy lost by the drop to its surroundings and $K_c$ is the thermal conductivity of air.   Assuming that the latent heat released by condensation dominates over changes in the sensible heat of the droplet, this energy is given by,
\begin{eqnarray}
4\pi a^2\rho L\frac{ d a }{dt} = \frac{dQ}{dt}
\label{eq:heat_phase_change}
\end{eqnarray}
Here, $L$ is the latent heat of evaporation. Combining equations (\ref{eq:condensation},\ref{eq:heat_conduction},\ref{eq:heat_phase_change}) and eliminating the $Q$ term we get,
\begin{eqnarray}
\frac{K_c}{D L}=\frac{\rho_{\infty}-\rho_{a}(T_a)}{T_a-T_{\infty}}
\label{eq:heat_balance}
\end{eqnarray}
Performing a Taylor's series expansion of $\rho_{a}$ about $T_{\infty}$ and neglecting $\mathcal{O}(T_a -T_{\infty})^2$ and higher terms we evaluate it at $T_a$. Combining it with equation (\ref{eq:heat_balance}) to eliminate $T_a$ results in,
\begin{eqnarray}
\rho_{a}(T_a) =  \rho_{a}(T_{\infty})+\Pi \left[ \rho_{\infty}-\rho_{a}(T_a)\right],  \\
\label{eq:heat_balance_taylor}
\end{eqnarray}
Here the dimensionless number $\Pi$ is defined as,
\begin{eqnarray}
&\Pi = \frac{DL}{K_c} \frac{d\rho_a}{dT}  \biggr|_{T_{\infty}} 
\label{eq:heat_balance_taylor_PI}
\end{eqnarray}
Since the Clausius–Clapeyron relation is a closed form expression it is straightforward to evaluate the density derivative with respect to temperature. Substituting $\rho_{a}(T_a)$ into equation (\ref{eq:condensation}) to remove the dependance on $T_a$ gives,
\begin{eqnarray}
\frac{da}{dt}=\frac{D}{\rho \, a [1+\Pi]}\left[\rho_{\infty}-\rho_{a}(T_{\infty})\right]
\label{eq:condensation_full}
\end{eqnarray}

Different regions of the air parcel can have differing $\rho_{\infty}$ due to large scale turbulent eddies acting on the mean water vapour gradient and generating fluctuations with Gaussian statistics\citep{kulmala1997effect,chandrakar2016aerosol}. This exchange can lead to excess water vapour in certain regions and so act as a stochastic source of supersaturation that can drive condensation. To model the non-deterministic water vapour exchange we discretise the air parcel into $N_p$ packets of equal volume $V$, each with its own collection of droplets. The water vapour content of each packet $m$ is given as, $w_m=\rho_{\infty,m}V$ and $w_m-<w>$, in the absence of condensation, is simulated as an Ornstein–Uhlenbeck process with a characteristic time for reversion of $\tau_E$, the Eulerian integral time scale, corresponding the largest turbulent eddies. The standard deviation of the process is a fixed fraction of the mean water vapour content $<w>$, which we shall denote as $f$. Typically, $f$ is $1\%$, but we will explore the effect of varying this parameter. Condensation acts as a sink for $w_m$ through droplets draining water vapour from the gas phase. We resolve this continuous evolution along with the stochastic behaviour, due to turbulent water vapour fluctuations, to accurately simulate the condensation growth of water droplets in turbulent clouds.

Similar to water vapour, droplets are also mixed by turbulence. To model this we move droplets from one packet to another to mimic movement from one history of water vapour fluctuation and local turbulence intensity to another. The rate at which this occurs set to be $1/\tau_E$, where $\tau_E$ is the integral time scale.

In contrast to continuous condensation driven droplet growth, growth due to collisions is discrete. Hence in our evolution study we track the discrete drop size distribution. However, there are a large number of droplets with a typical drop number density of $10^8$ per $m^3$\citep{li2017eulerian}. It is beyond the scope of our study to track the trajectory of these droplets within the large volumes swept out by the air parcel as it rises from the ground. Instead we track collisional growth through Monte Carlo simulation of the rate law. Binning multiple real droplets into a Monte Carlo droplet allows for increased computational efficiency while still retaining accuracy and the discrete nature of the distribution.

The rate of change of droplet concentration follows a two-species rate law for collisional growth given as,
\begin{eqnarray}
\frac{dn_i}{dt}= -C_{ij} n_in_j
\label{eq:coll_form1}
\end{eqnarray}
where $n_i$ corresponds to the number density of species $i$ in the bulk. Only two species interaction is considered due to the dilute nature of the cloud system as, for example, \cite{grabowski2013growth} suggest that the drop volume fraction is $\textit{O}(10^{-6})$ . $C_{ij}$ is the rate constant set by the turbulent flow, gravity, non-continuum hydrodynamic interactions and inertial clustering.

Within a packet there are many droplet pairs that could potentially collide and coalesce. In fact, in a given packet with $N$ Monte Carlo droplets each of which represents $N_d$ real droplets, $N[N-1]/2$ Monte Carlo droplet collisions are possible.   If we tested every one of these pairs in a given time step $dt$, the appropriate probability for droplet collision to reproduce the kinetic rate of the actual drops would be,
\begin{eqnarray}
P^*= \frac{C_{ij} N_{d,j} dt}{V}
\label{eq:prob_Monte_Carlo}
\end{eqnarray}
for each Monte Carlo pair with $N_{d,i} \leq N_{d,j}$.
Due to computational constraints, however, we will only test, at any given time step, $ \mathcal{G} (N/2)$ randomly chosen pairs within a packet for collisions, where $\mathcal{G}$ is the greatest integer function. Thus, the new probability of collision $P$ of any chosen Monte Carlo droplet pair is given as,
\begin{eqnarray}
P= \frac{1}{ \mathcal{G} (N/2)}\frac{N[N-1]}{2}\frac{C_{ij} N_{d,j} dt}{V}
\label{eq:prob_Monte_Carlo_evolution}
\end{eqnarray}
If this $P$ is greater than a random number chosen with equal probability between 0 and 1, collision and subsequent coalescence occurs. In this case the radii of the Monte Carlo droplet $i$ and $N_{d,i}$ of the droplets represented by Monte Carlo droplet $j$ are updated. If $N_{d,i} = N_{d,j}$ they are assigned $\mathcal{G} (N_{d,i}/2)$ and $\mathcal{G} (N_{d,i}/2)+1$ droplets respectively, along with the updated radii for each of them.

To calculate $P$ in equation (\ref{eq:prob_Monte_Carlo_evolution}) information concerning the rate constant $C_{ij}$ is needed. To evaluate this we consider the nature of the collisions including the role droplet inertia plays. A measure of droplet inertia is the non-dimensional Stokes number, given as $St= \tau_{p}/\tau_{f}$. Here, the particle response time is $\tau_{p}=2a^2 g/[9\nu] $, with $g$ the acceleration due to gravity, and $\tau_{f}$ the characteristic fluid time, which for the sub-Kolmogorov sized droplets in turbulent clouds is the Kolmogorov time scale. The mean Stokes number, relevant to collision of droplets with disparate sizes, is found to be significant, even reaching $\textit{O}(1)$ values under some conditions in clouds\citep{ayala2008effectsa}. This will enhance the local concentration and increase the frequency of collisions. To account for this effect at low to moderate Stokes number, it is possible to describe the rate constant as a product of the collision kernel and the radial distribution function at contact\citep{chun2005clustering,ireland2016aeffect,dhanasekaran2022model}. This is given as,
\begin{eqnarray}
C_{ij}=g_{ij}(r)|_{r=a_i+a_j}\,K_{ij}
\label{eq:rate_constant}
\end{eqnarray}
Here $K_{ij}$ is evaluated for zero droplet inertia and $g_{ij}(r)$ is the radial distribution function, capturing local concentration enhancement due to droplet inertia for droplets of radii $a_i$ and $a_j$ with a separation $r$. Inertia can potentially alter $K_{ij}$ through its role in the trajectory of the colliding droplets. Studies that account for this mechanism are limited. \cite{davis1984rate} considers it for sedimenting particles but does not include turbulence or a comprehensive treatment of non-continuum hydrodynamics. The DNS study by \cite{li2018effect} while incorporating the role of particle inertia does not include a proper treatment of the hydrodynamic interactions.\cite{chen2018turbulence} included long-range continuum hydrodynamic interaction and did not incorporate particle inertia. Since we find that an accurate treatment of hydrodynamic interactions is important we use the results by \cite{dhanasekaran2021collision_a} and consider droplet inertia through preferential concentration alone.

The rate constant $K_{ij}$ can be expressed as the product of an ideal collision rate obtained by neglecting hydrodynamic and colloidal interactions between the drops and a collision efficiency which represents the change in collision rate due to such interactions.
In the absence of droplet inertia, the ideal collision rate for pure differential sedimentation was calculated by \cite{smoluchowski1918versuch} and shown to be sensitive to difference in size. The collision rate of drops due to turbulent shearing motions in the absence of sedimentation was derived by \cite{saffman1956collision} but it did not include the dependence on the Reynolds number based on the Taylor microscale of turbulence $\textrm{Re}_{\lambda}$. Variation with this parameter has been studied by \cite{dhanasekaran2021collision_a} who find a $20\%$ change in collision rate between $\textrm{Re}_{\lambda}=90$, typical in DNS studies, and the more realistic value of 2500 encountered in turbulent clouds. Both turbulence and differential sedimentation are expected to be important for collisions in the 'size-gap' and this coupling has been resolved in their collision rate study. Since the droplets are present in air, hydrodynamic interactions incorporating the breakdown of continuum upon close approach of the droplets are expected to strongly influence collision events, while van der Waals effects and droplet deformation are not expected to play an important role\citep{sundararajakumar1996non}. While a few studies have calculated collision rates with non-continuum effects\citep[see][]{davis1984rate,chun2005coagulation} the most comprehensive analysis is by \cite{dhanasekaran2021collision_a}. They calculate the retardation due to hydrodynamic interactions and report the collision efficiency ($\beta$) as a function of the relative size of the interacting droplets, the Knudsen number Kn=$2\lambda_g/[a_1+a_2]$ (where $\lambda_g$ is the mean free path of air), and the relative strength of differential sedimentation velocity and the turbulent shear. Although we assume viscous forces dominate in determining the hydrodynamic interactions between drops, the terminal sedimentation velocities of each drop account for non-linear drag using correlations given in table 5.2 of \cite{clift2005bubbles}. The turbulent velocity depends on the local shear rate $\Gamma_{0}$. The probability distribution function of local Kolmogorov properties exhibits large tails, indicating increased likelihood of intermittent shear rates relative to a normal distribution. $\Gamma_{0}$ was modelled by \cite{koch2002coagulation} as a log-normal distribution, with the constitutive Gaussian statistics obtained from an Ornstein–Uhlenbeck process dependent on both the Kolmogorov and integral scales of turbulence. We take this result and create a unique realisation of the turbulent shear rate in each packet. Thus, in addition to resolving intermittent collisions, we also capture multiple realisations of turbulence.


Droplet inertia creates a delay in response to the background turbulence leading to clustering in certain regions of the flow. In addition to $St$ based on individual droplet motion, it depends on the strength of differential sedimentation, captured through the difference in the settling parameters of the two drops $\Delta S_{v,ij}=|\tau_{p,i}-\tau_{p,j}|g/u_{\eta}$. Here the Kolmogorov velocity $u_{\eta}$  is given as $[\nu\epsilon]^{1/4}$, where $\epsilon$ is the turbulent dissipation rate and $\nu$ is the kinematic viscosity of the gas. $St=\textit{O}(1)$ leads to maximum inertial clustering. However DNS studies by \cite{ayala2008effectsa} and \cite{dhariwal2018small} showed that decorrelation through differential sedimentation occurs rapidly as the particle pair sizes become different, indicating $g_{ij}(r)$ is largest for $\Delta S_{v,ij} \ll 1$. All of these effects have been captured and $g_{ij}(r)$ validated against DNS results in the inertial clustering model developed by \cite{dhanasekaran2022model}. This will be used in our evolution study.

The mean-free path, $\lambda_0$, and kinematic viscosity, $\nu$, are sensitive to temperature and pressure. These thermodynamic quantities evolve as the cloud packet rises and the gas expands and cools and as the condensation reduces the water vapour pressure and the gas temperature. To determine the corresponding instantaneous viscosity we use the Sutherland viscosity law that is a function of temperature\citep{sutherland1893lii}. To obtain the instantaneous mean free path we use the expression given by \cite{jennings1988mean} that is a function of $\nu$, temperature, and pressure. We evolve $\lambda_0$ and $\nu$ in this manner unless other conditions are explicitly stated.

\section{Results and discussion}\label{sec:results}

Due to the wide range of parameters under consideration we will use typical values for the turbulent and microphysical parameters unless otherwise stated. Their default values are $\epsilon=0.01m^2/s^3$, $\textrm{Re}_{\lambda}=2500$, $f=1\%$, ground temperature $T_0$ of 293 K, ground pressure $P_0$ of 1 bar and  up-draft velocity of 1 m/s. The mean-free path and kinematic viscosity, $\lambda_0$ and $\nu$, evolve accordingly unless otherwise explicitly stated. The default simulation parameters are: $V=1 m^3$, $N_p=100$, initial Monte Carlo droplets per packet of 100, and an initial droplet radius distribution that is uniform random between 0.1 to 1 $\mu m$.

\begin{figure}
\hspace*{-0cm}
\includegraphics[scale=0.3]{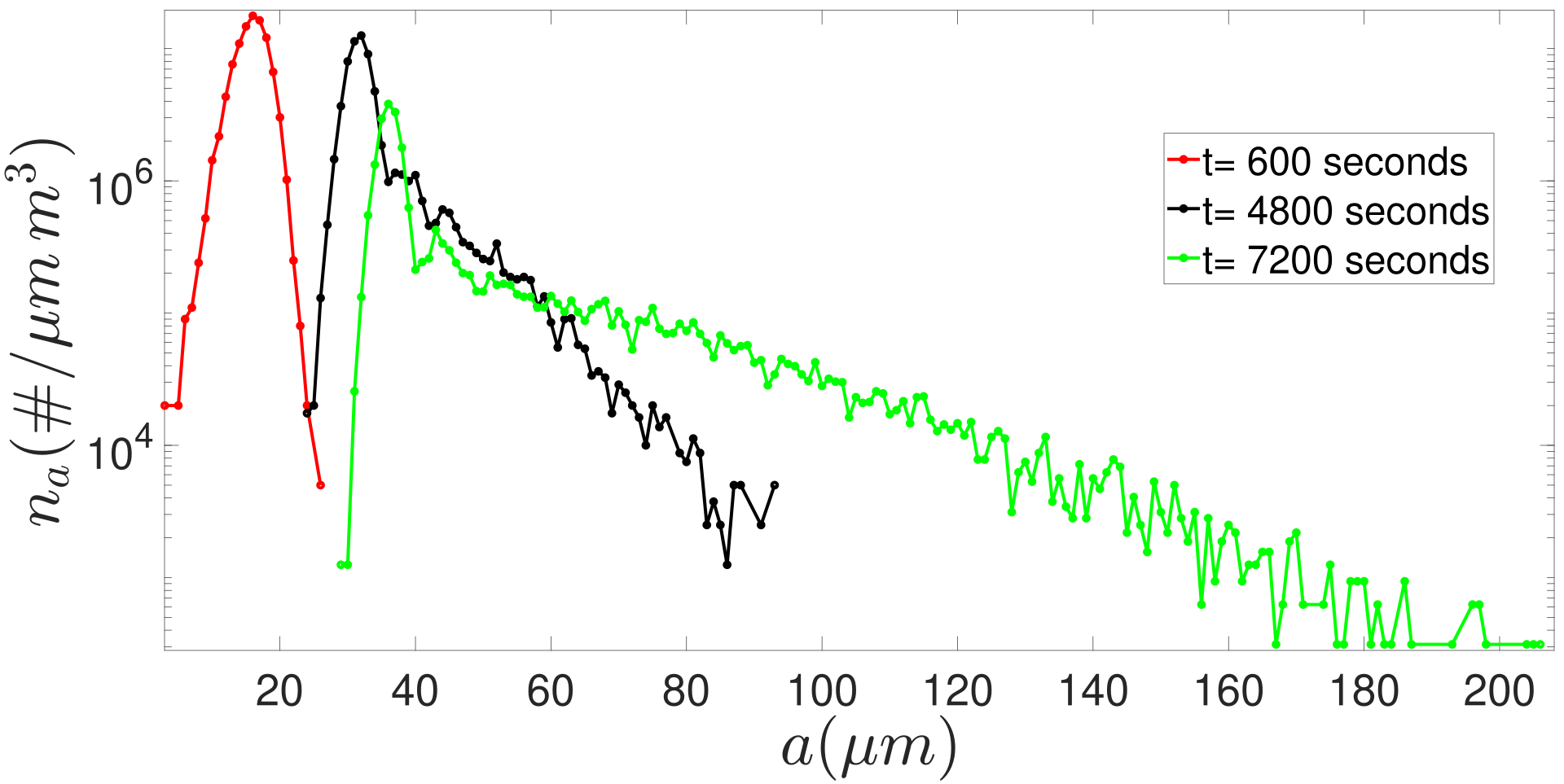}
\caption{The number  of droplets per unit volume with radii between $a$ and $a+da$, which is  denoted as $n_a$, is plotted against radius during the early (600 seconds), intermediate (4800 seconds) and long (7200 seconds) time evolution of our model. The drop size distribution does not conform to any specific form at all times. Starting off Gaussian it transitions to a skewed distribution with a long tail at large drop sizes as the system evolves.}
\label{fig:distribution}
\end{figure}

The drop size distribution $n_a$ defined as the number of drops per unit volume of space per unit radius is plotted for this default case in figure \ref{fig:distribution} at 600, 4800 and 7200 seconds as a function of the drop radius.  At early times, the drop size distribution evolves by condensation into a nearly Gaussian distribution because the water vapour fluctuations are assumed to have a normal spread. As the drops grow, collisions begin to dominate the growth of still larger droplets. This process is not Gaussian and the size distribution forms a tail of large drops.  Collisions driven by the large drops in the tail drain drops from the Gaussian portion of the distribution further lengthening the tail.  The long tails appears to exhibit exponential behaviour, which is in line with predictions for sedimentation dominated collisional growth \citep{van1988scaling,westbrook2004theory}.

While the full drop size distribution can provide useful insight it is not feasible to report it at every instant for all the simulations performed. Instead we will report two metrics of the distribution, namely the volume averaged mean radius $a_v$ and the dispersion $\mathcal{D}$.  Here, $a_v$ is a measure of the size of drops as seen by the liquid water content and is given as,
\begin{eqnarray}
a_v=\frac{\sum a_i^4}{\sum a_i^3}
\label{eq:vol_avg_rag}
\end{eqnarray}
$\mathcal{D}$ is a non-dimensional estimate of the spread of the distribution and is given as,
\begin{eqnarray}
\mathcal{D}=\frac{ \sigma_a}{<a_i>}
\label{eq:vol_std_rag}
\end{eqnarray}
where, $\sigma_a$ is the standard deviation and $<a_i>$ is the number averaged radius of the drops.

\begin{figure}
\hspace*{-0cm}
\includegraphics[scale=0.3]{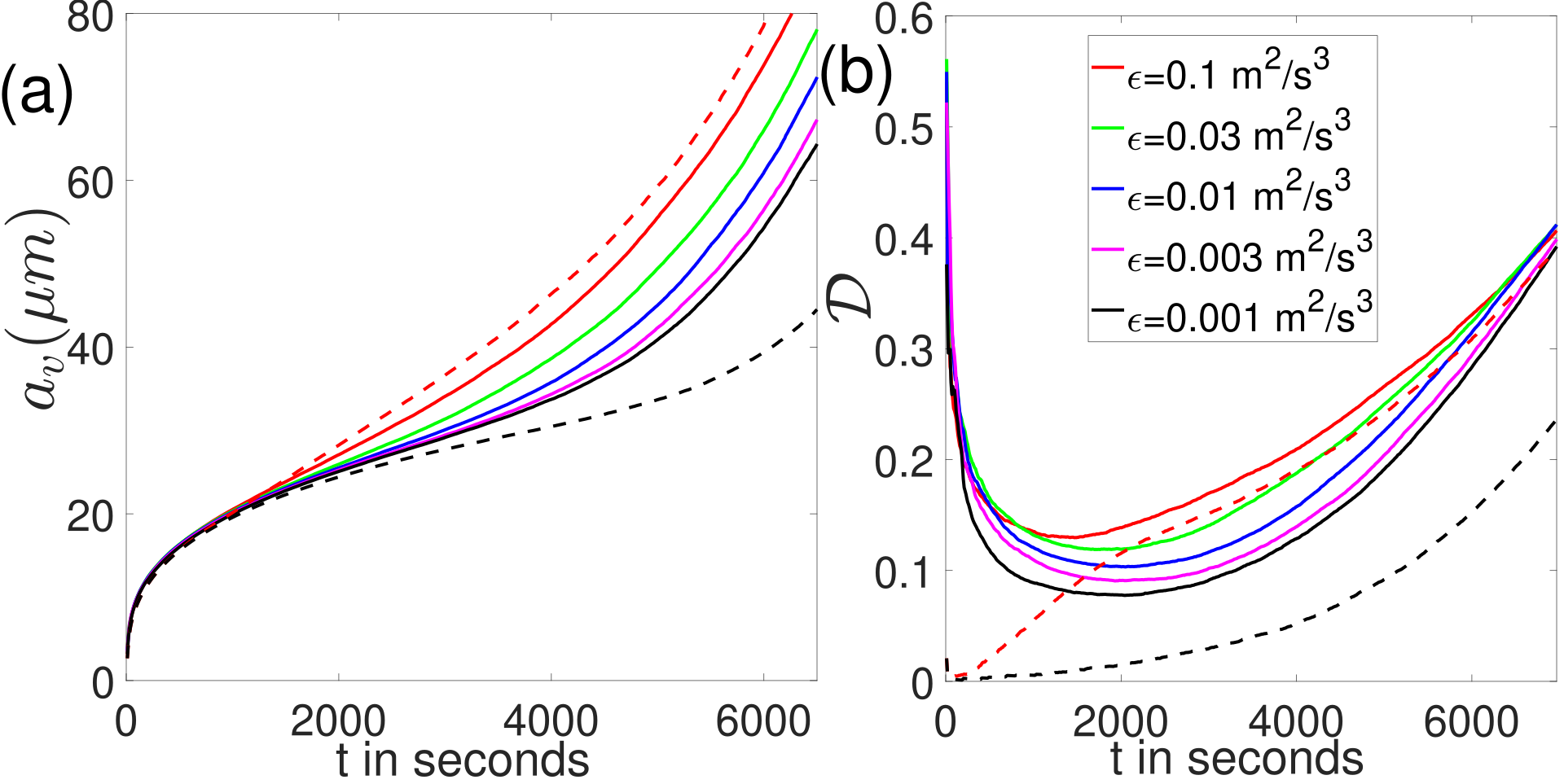}
\caption{Evolution of (a) the volume averaged mean radius $a_v$ and (b) the dispersion $\mathcal{D}$ as drops cross the 'size-gap'.   The solid lines are obtained from simulations including all of the mechanisms in our model and correspond to various turbulent dissipation rates $\epsilon$ noted in the legend.  The dashed lines are from simulations with condensation and coupled turbulent-gravitational collisions but without water vapour fluctuations and turbulent mixing.  The upper dashed curve corresponds to $\epsilon=0.1m^2/ s^3$ while the lower is for $\epsilon=0.001 m^2/ s^3$. }
\label{fig:eps_wv01_Var_evol}
\end{figure}

Using the metrics in equations \ref{eq:vol_avg_rag} and \ref{eq:vol_std_rag} the drop size evolution as a function of time is shown in figure \ref{fig:eps_wv01_Var_evol}. As expected from figure \ref{fig:distribution} the radius $a_v$ monotonically increases while $\mathcal{D}$ and the growth rate of the radius reach minima at intermediate times. The period of slower changes in radii, occurring after about an hour of evolution, comes when $a_v$ lies between 15 and 40 $\mu m$ and droplets struggle to make it through the size-gap. Comparing figure \ref{fig:eps_wv01_Var_evol} (b) and figure \ref{fig:eps_wv01_Var_evol} (a) shows that the minimum of $\mathcal{D}$ occurs while $a_v$ lies squarely within the size-gap.  A wide variety of mechanisms of droplet growth influence the evolution of   $a_v$ and $\mathcal{D}$ as the drops approach, enter and leave the size-gap region.  To gain physical insight into the influence of these different mechanisms, it will be useful to sequentially turn off different growth mechanisms and observe the resulting drop size evolution.  Using the understanding of physical mechanisms obtained in this way, we will then present results for the full model at different atmospheric conditions. 

\begin{figure}
\hspace*{-0cm}
\includegraphics[scale=0.3]{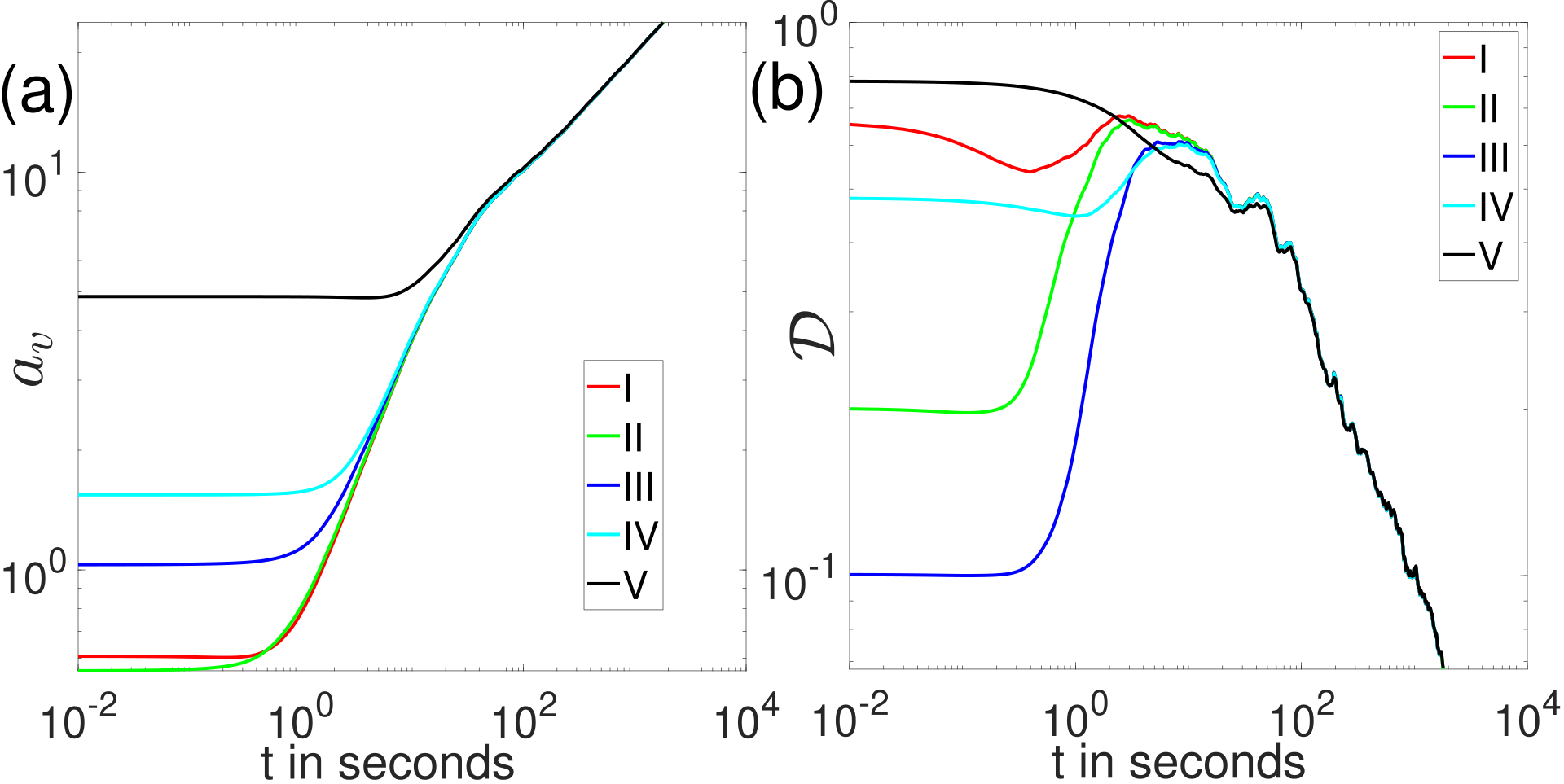}
\caption{Droplet growth by condensation alone in a rising cloud packet to a state that is independent of the initial conditions.  Figure (a) gives the volume averaged mean radius $a_v$ and (b) the dispersion $\mathcal{D}$.  The water vapour fluctuations are $f=1\%$.  In the Legend 'I','II','III','IV', and 'V' correspond initial Gaussian drop size distributions with means of 0.25, 0.5, 1, 1, 2 $\mu$m and standard deviations of 0.25, 0.1, 0.1, 0.5, 2 $\mu$m, respectively. }
\label{fig:cond_only_evol}
\end{figure}

Condensation dominates the early evolution and produces a drop size distribution that is independent of the initial distribution of drops with radii on the order of microns. This occurs because condensation favors the growth of smaller droplets relative to larger ones leading to monodisperse drops when the water vapor content is uniform or to a distribution that mirrors the distribution of water vapor content. To demonstrate this behavior, we consider the growth of drops resulting from condensation alone, i.e., no collisions or droplet mixing, as the cloud packet rises. This is performed with water vapor fluctuations of $f=1\%$ and mixing of water vapor between the different packets. Figure \ref{fig:cond_only_evol} shows the time variation of the mean radius and dispersion for initially Gaussian drop size distributions, with the various means and standard deviations noted in the figure caption.  All the cases collapse in a matter of minutes to a state that is independent of initial conditions at radii for which the growth in the full model would still be dominated by condensation. Thus, in the balance of our study, we consider sub-micron sized cloud condensation nuclei and do not further explore the effect of the initial drop size distribution.  The dispersion of droplet size in Figure  \ref{fig:cond_only_evol} (b) decreases at long times because the transfer of water vapor to the liquid phase attenuates the water vapor fluctuations.

The rate of growth of drops by condensation decreases with increasing droplet size and, in addition, the water vapor content in the packet decreases as a larger portion of the moisture is passed into the liquid phase.  As a result the drop radius scales as $t^{1/2}$ and the time rate of change of the droplet radius becomes exceedingly small within the 'size-gap'.  Incorporation of collisions driven purely by differential sedimentation is not sufficient to allow droplets to grow through the size gap in a reasonable time period.  Condensation driven growth in a gas with uniform water vapor fluctuations leads to perfectly monodisperse drops and even in the presence of water vapor fluctuations the dispersion decreases with time as illustrated in  Figure  \ref{fig:cond_only_evol} (b). 

\begin{figure}
\hspace*{-0cm}
\includegraphics[scale=0.3]{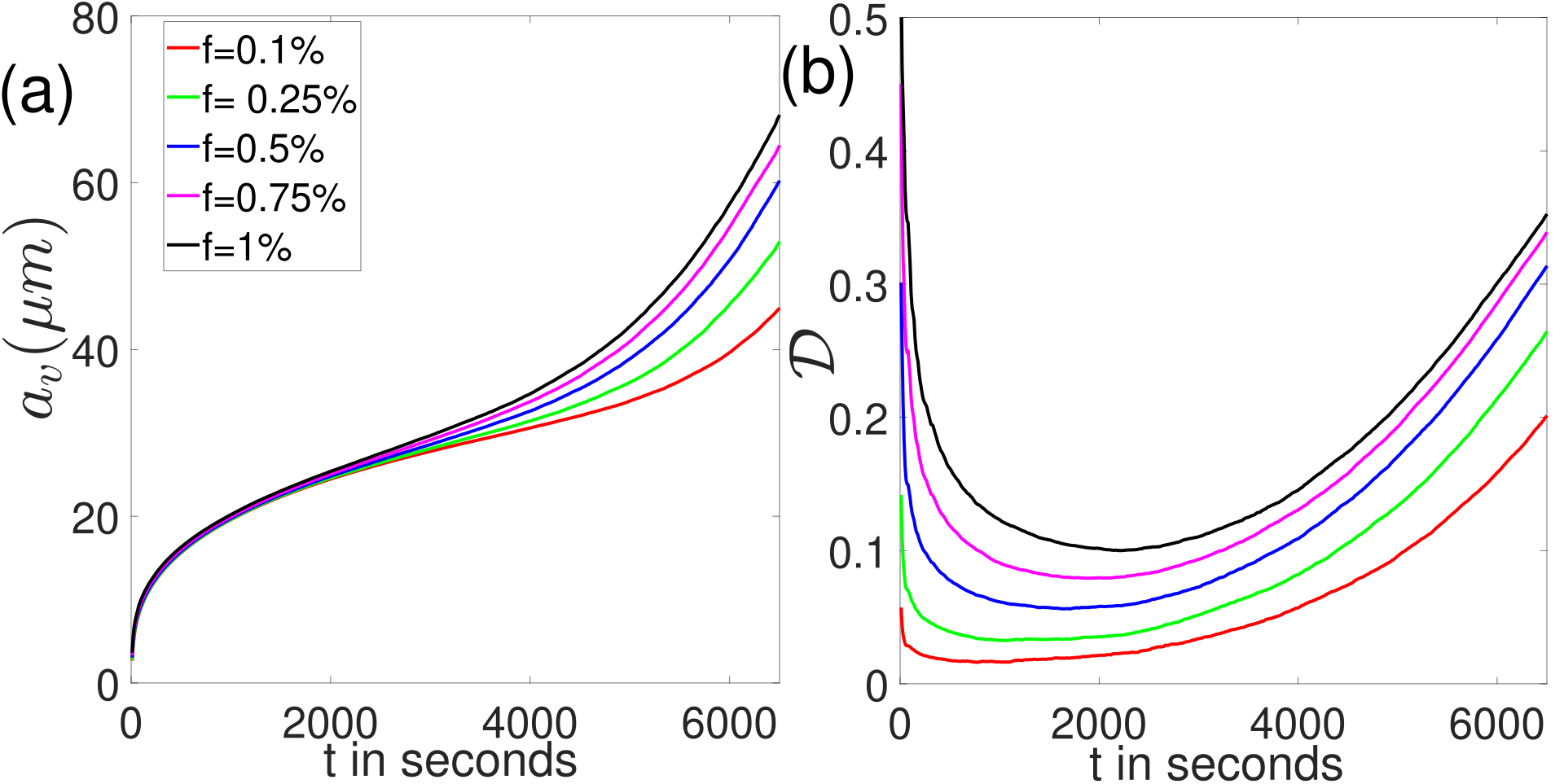}
\caption{Droplet growth by condensation and gravity-driven collisions in the presence of water vapor fluctuations and turbulent mixing of droplets but without turbulence-induced collisions. The volume weighted mean radius $a_v$ (a) and the dispersion $\mathcal{D}$ (b) are shown as a function of time.  Increasing $f$ from $0.1\%$ to $1\%$ enhances the growth rate.}
\label{fig:fluc_Var_evol}
\end{figure}

To assess the ability of condensation with water vapor fluctuations and differential sedimentation driven collisions acting alone to allow drops to cross the size-gap region, we present simulations including these effects but excluding turbulence driven collisions in Figure  \ref{fig:fluc_Var_evol}.  Initially $\mathcal{D}$ decreases due to the faster growth of smaller droplets by condensation. However, the polydispersity reaches a plateau resulting from the variation of water vapour partial pressure which continues until the droplets are moderately large and sedimentation driven collisions between large drops with modest size differences are able to enhance polydispersity and drive faster growth.  Thus, droplet growth in the absence of turbulence-driven collisions slows significantly in the size gap but eventually gravity-driven coalescence produces larger drops after a time delay that decreases with  increasing water vapor variation $f$.


\begin{figure}
\hspace*{-0cm}
\includegraphics[scale=0.3]{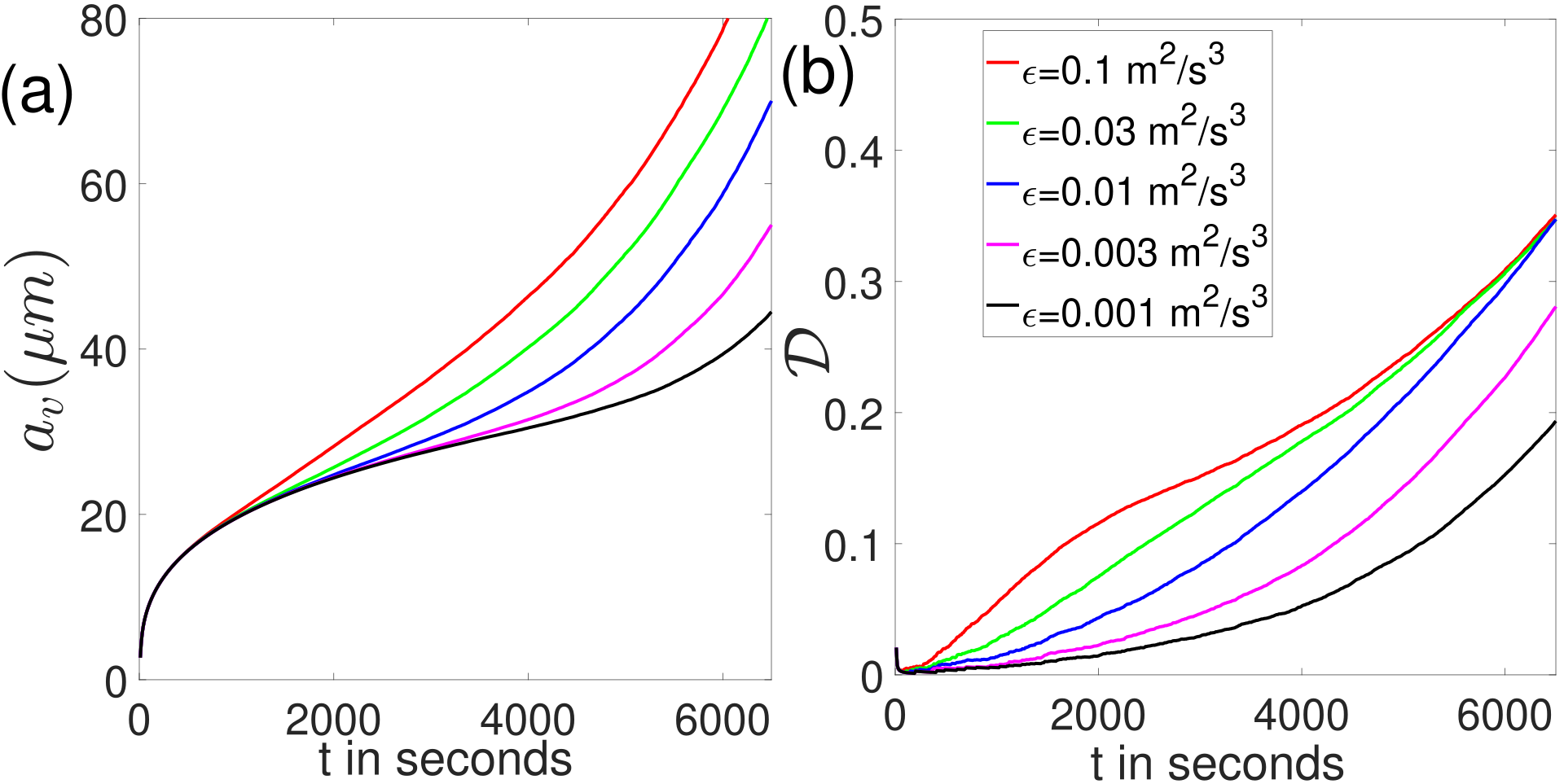}
\caption{Droplet growth by coalescence resulting from the coupled effects of turbulent shear and differential sedimentation along with condensation but in the absence of water vapor fluctuations and turbulent mixing of droplets.  
 Figure (a) and (b) show the evolution of $a_v$ and  $\mathcal{D}$ for turbulent dissipation rates $\epsilon$ ranging from $0.001$ to $0.1 m^2/ s^3$.}
\label{fig:eps_wv0_Var_evol}
\end{figure}

Turbulent shearing motion can drive collisions between drops of equal size and so they provide an alternative mechanism to overcome the monodisperse nature of the drop size distribution produced by the early condensation-driven growth even in the absence of fluctuations in water vapor content.  In figure \ref{fig:eps_wv0_Var_evol} we illustrate the drop growth that occurs by condensation and coalescence driven by the coupled effects turbulence and differential sedimentation in the absence of water vapour fluctuations and turbulent mixing of droplets. We include the influence of inertial clustering of droplets on the collision frequency.   To assess the role of the turbulence level of the cloud, results are included for a range of turbulent dissipation rates. In the early evolution $\mathcal{D}$ is small for all $\epsilon$ as there is no polydispersity generating mechanism and condensation favours a nearly monodisperse distribution. When the drops are about $20 \mu m$, turbulent collisions start to takeover. The rate at which the 'size-gap' is crossed increases as the turbulent dissipation rate and thus the Kolmogorov shear rate $\Gamma_0$ driving droplet collisions increases. This result is consistent with the large-eddy simulations with Lagrangian microphysics scheme carried out by \citep{chandrakar2024turbulence}, which showed that inclusion of a simpler model of the turbulent collision kernel than that provided here enhances droplet growth and brings it closer to experimental observations. 

Having seen that either turbulent-shear driven collisions or water vapor fluctuations acting individually can limit the decrease in the polydispersity of the drops and allow growth of the drops to sizes at which differential-sedimentation-driven coalescence produces more rapid growth, we now consider the evolution of droplet size including all features of the model as solid lines in figure \ref{fig:eps_wv01_Var_evol} for several turbulent dissipation rates and $f=1\%$.  As expected, turbulent shear limits the decrease in polydispersity for drops entering the size-gap region and higher turbulence leads to more rapid growth of the drops' volume-averaged radius.  The dashed curves in figure \ref{fig:eps_wv01_Var_evol} show simulation results without water vapor fluctuations and droplet mixing for the lowest and highest dissipation rates.  As expected, at low turbulence levels ($\epsilon = 0.001 m^2/s^3$), water vapor fluctuations play a crucial role in maintaining finite drop size dispersion and allowing growth through the size-gap region, whereas the calculations without water vapor fluctuations still leaves $a_v$ in the size-gap at $t=6000 s$.  

More surprisingly, however, the presence of water vapor fluctuations increases the time spent in the size-gap region at high turbulent dissipation rates ($\epsilon = 0.1 m^2/s^3$).
 In the early, condensation-controlled evolution, the growth rate of the drop radius is higher in the more polydisperse drop distribution occurring in the presence of water vapour fluctuations and turbulent droplet mixing.  Later, however, water-vapor-fluctuation-driven polydispersity hinders the growth of drops by turbulence-driven collisions. This occurs despite the fact that the ideal turbulent collision rate, $n_1n_2\Gamma_0[a_1+a_2]^3$, is insensitive to the difference in the size of colliding droplets. The dependence of the drop coalescence rate  on $a_2/a_1$ arises instead from the non-continuum hydrodynamic interactions, which result in a collision efficiency that is $30\%$ smaller at $a_2/a_1=0.4$ than at $a_2/a_1=0.9$ \citep{dhanasekaran2021collision_a}, and from inertial clustering, which is less effective in enhancing collision frequency as  polydispersity increases. Thus, coalescence dominated by turbulent shear is hindered by polydispersity. It should be noted that collision efficiency for differential sedimentation also decreases (by $50\%$) when $a_2/a_1$ is decreased from 0.9 to 0.4 \citep{dhanasekaran2021collision_a}. However, the dependence of the ideal rate, $n_1n_2 \frac{2}{9}\rho g[a_2^2-a_1^2] [a_1+a_2]^2/\mu $, on the size ratio more than makes up for this deficit.

\begin{figure}
\hspace*{-0cm}
\includegraphics[scale=0.3]{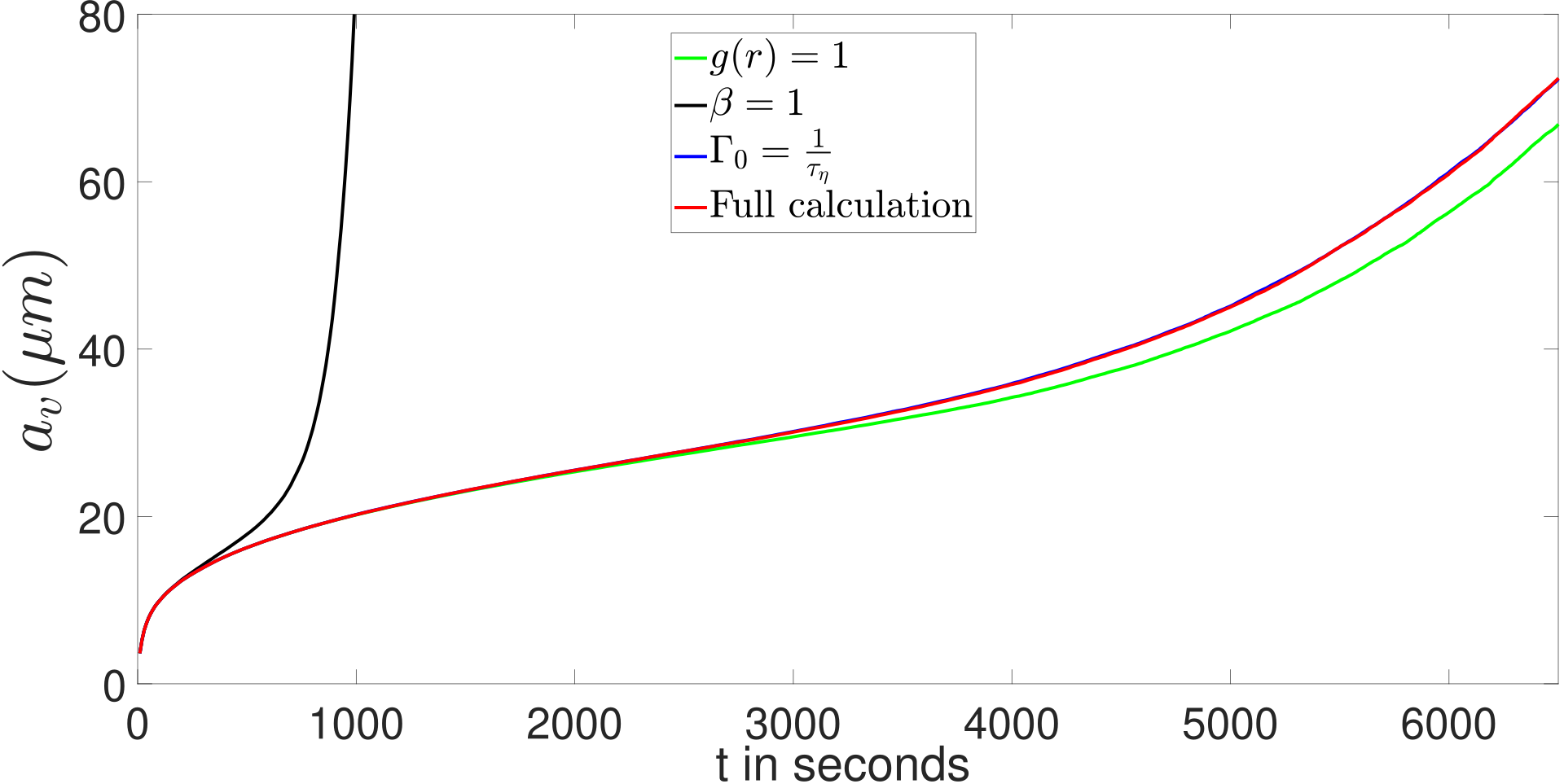}
\caption{The temporal evolution of the volume averaged mean radius $a_v$ for full model is compared with cases in which hydrodynamic interactions are neglected so that the collision efficiency $\beta=1$, in which inertial clustering is neglected so that the pair distribution function $g(r)=1$, and in which fluctuations in the turbulent shear rate are neglected so that $\Gamma_0=\frac{1}{\tau_{\eta}}$.}
\label{fig:factors_combined}
\end{figure}

The foregoing discussion demonstrated that turbulent-shear-induced collisions along with water vapor fluctuations play a critical role in allowing droplets to grow to sizes where sedimentation-induced growth takes over.  Two other features of turbulence that have been proposed to accelerate droplet growth are inertial clustering of droplets (measured by $g(r)$)  and the variability of the turbulent shear rate ($\Gamma_0$) due to intermittency.  Figure \ref{fig:factors_combined} illustrates the importance of these factors in our model in comparison with the effects of non-continuum hydrodynamic interactions captured by the collision efficiency $\beta$. The figure compares the full calculation with a calculation that neglects hydrodynamic interactions $\beta=1$, a simulation with no fluctuations in the local Kolmogorov shear rate ($\Gamma_0=1/\tau_{\eta}$), and one that neglects inertial clustering $g(r)=1$. Clearly the common practice of neglecting droplet interactions by setting the collision efficiency equal to one greatly overestimates the drop growth. In contrast, inertial clustering causes a modest enhancement of the growth rate and variations of the turbulent shear rate driving collisions have a much weaker impact.

Turbulent statistics are typically characterised by fat tails of the velocity-gradient probability distribution function. Using this idea \cite{kostinski2005fluctuations} proposed that a few 'lucky' droplets can collide due to abnormally large local turbulent shear rates and thus cross the 'size-gap'. However, our calculations in figure \ref{fig:factors_combined} show that these 'lucky' droplets do not grow large enough to significantly influence the evolution of the whole distribution. In fact the formation of moderately large 'lucky' drops simply enhances the polydispersity-driven retardation in growth at the higher dissipation rate that we discussed previously (not shown for sake of brevity).

\begin{figure}
\hspace*{-0cm}
\includegraphics[scale=0.3]{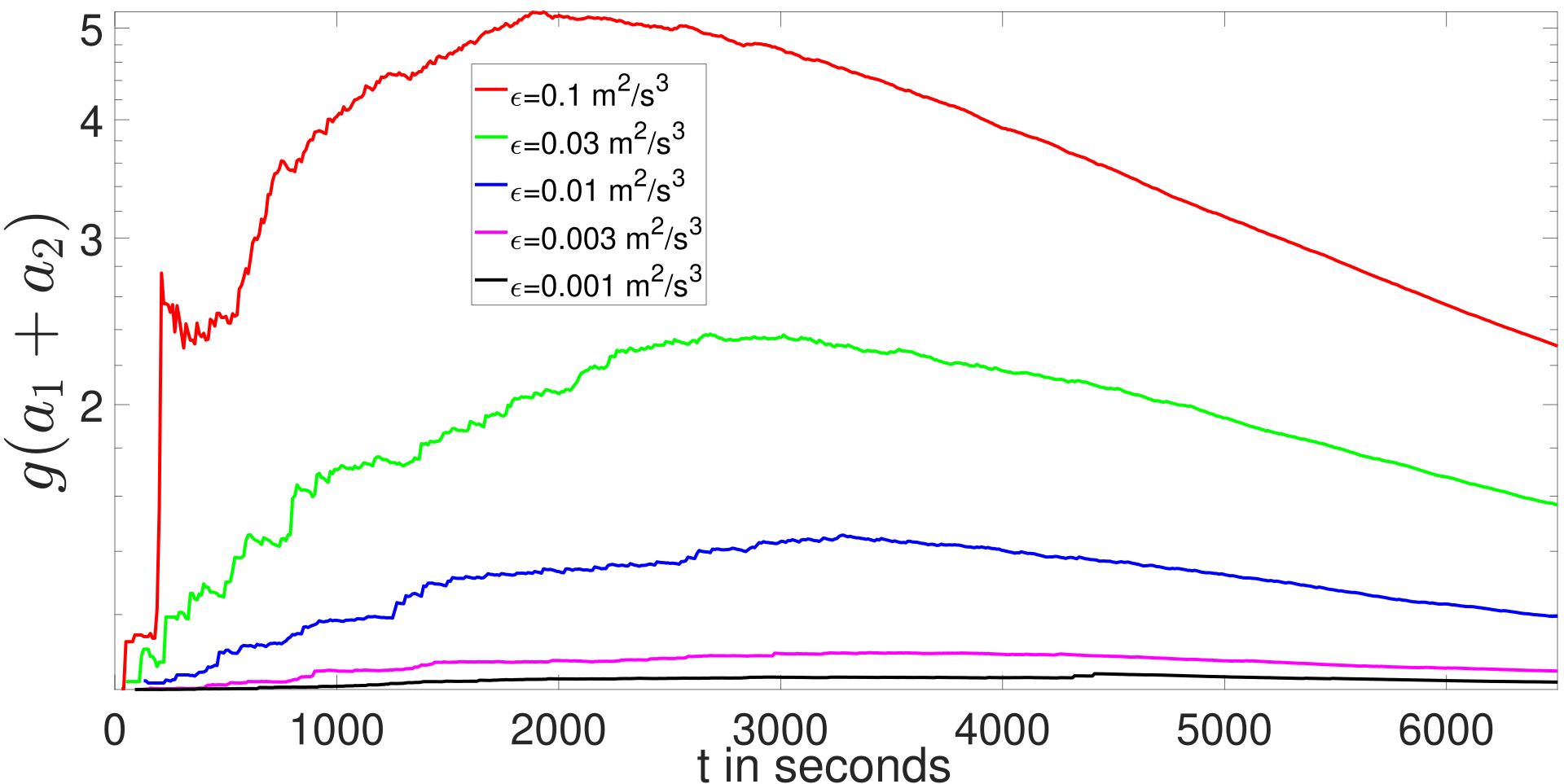}
\caption{The inertial-clustering enhanced pair distribution function $g(a_1+a_2)$ averaged over all colliding particles for the full calculation is shown to have a non-monotonic dependence on time with increasing peak values as the turbulent dissipation rate is increased for $\epsilon=0.001,0.003,0.01,0.03,0.1$ m$^2/$ s$^3$. }
\label{fig:gr_eps_wv01_Var_evol}
\end{figure}

\begin{figure}
\hspace*{-0cm}
\includegraphics[scale=0.3]{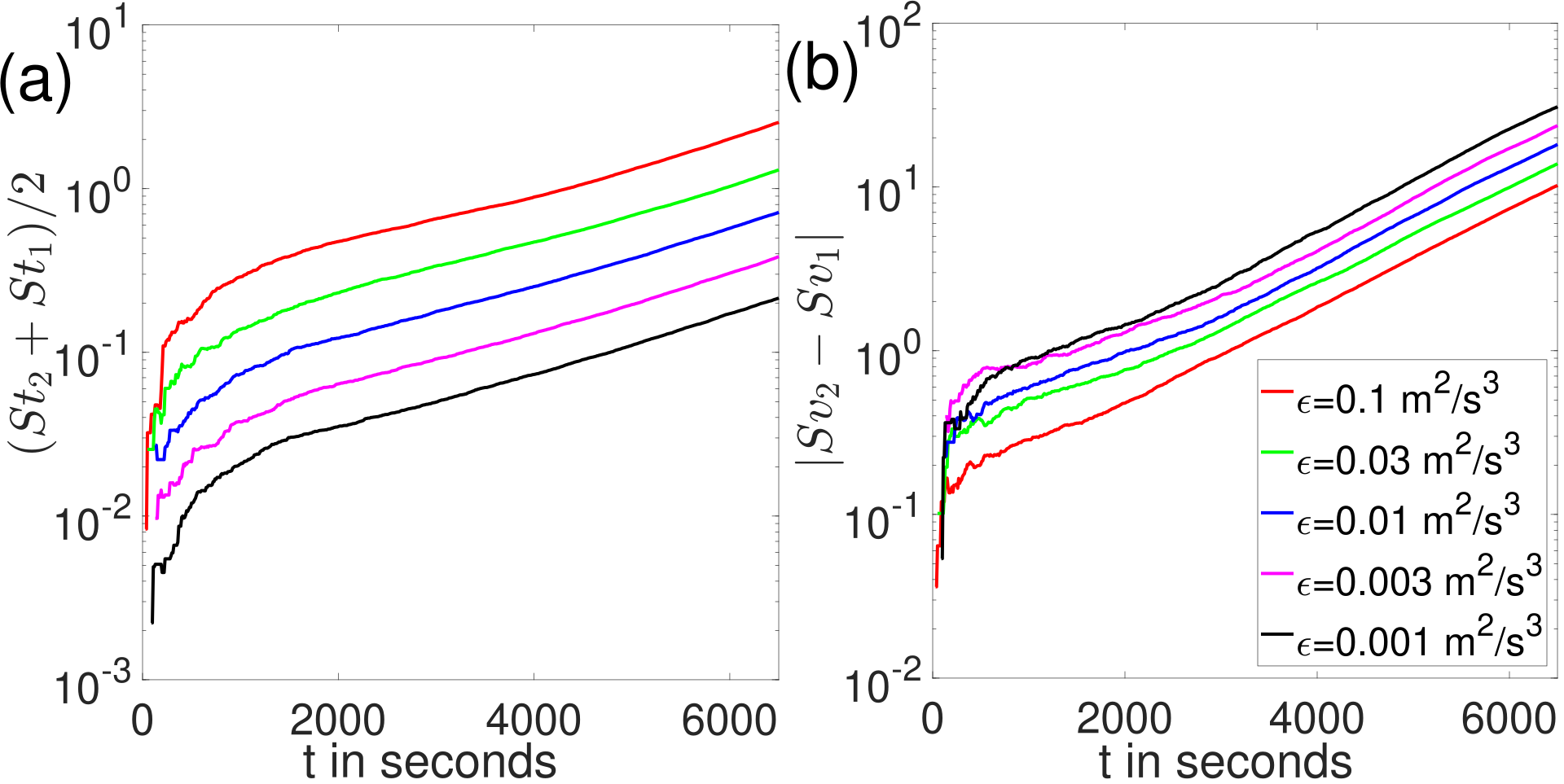}
\caption{Evolution of the two important parameters of inertial clustering as a function of time shown for $\epsilon=0.001,0.003,0.01,0.03,0.1$ m$^2/$ s$^3$. Figure (a) shows the mean Stokes number and figure (b) the magnitude of the difference in non-dimensional settling velocity of pairs of colliding particles.}
\label{fig:gr_stsv_eps_wv01_Var_evol}
\end{figure}

Inertial clustering has a more pronounced effect on the drop size evolution. Yet it does not lead to an order of magnitude enhancement of growth rate (see figure \ref{fig:factors_combined}). To understand this, we plot the value of the pair distribution function of contacting drops $g(a_1+a_2)$ averaged over all the successful collision events. This is shown in figure \ref{fig:gr_eps_wv01_Var_evol} for the full calculation evolution at all the values of the dissipation rate chosen in figure \ref{fig:eps_wv01_Var_evol}. A monotonic behaviour is not observed, instead a peak appears in the inertial clustering enhancement at about 1 hour of simulated rise of the cloud packet. To obtain insight into the peak and its modest value we consider the mean $St$ and the difference in $S_v$ of the colliding pairs. These should be $\textit{O}(1)$ and negligible respectively to have maximum enhancement of collision frequency via inertial clustering. The averages of these two parameters over the collisions occurring at a given time are shown in figure \ref{fig:gr_stsv_eps_wv01_Var_evol}. The Stokes number $St$ is shown in figure \ref{fig:gr_stsv_eps_wv01_Var_evol} (a) to monotonically increase with time across the parameter space. However, it does not reach  $\textit{O}(1)$ values for the lower dissipation rates before $\Delta S_v$ is  $\textit{O}(1)$  (figure \ref{fig:gr_stsv_eps_wv01_Var_evol} (b)). When turbulent collisions are dominant, as we have seen for $\epsilon=0.1 m^2/s^3$, inertial clustering can reduce the time to reach a mean radius of $40 \mu m$ by about $15\%$. This is line with the significantly large peak of $g(a_1+a_2)$ at $t=2000$s for $\epsilon=0.1 m^2/s^3$ observed in figure \ref{fig:gr_stsv_eps_wv01_Var_evol} (a).  This occurs because $St$ has increased to values around 0.5 where inertial effects are significant, while $\Delta S_v$ is still only about 0.5.

\begin{figure}
\hspace*{-0cm}
\includegraphics[scale=0.3]{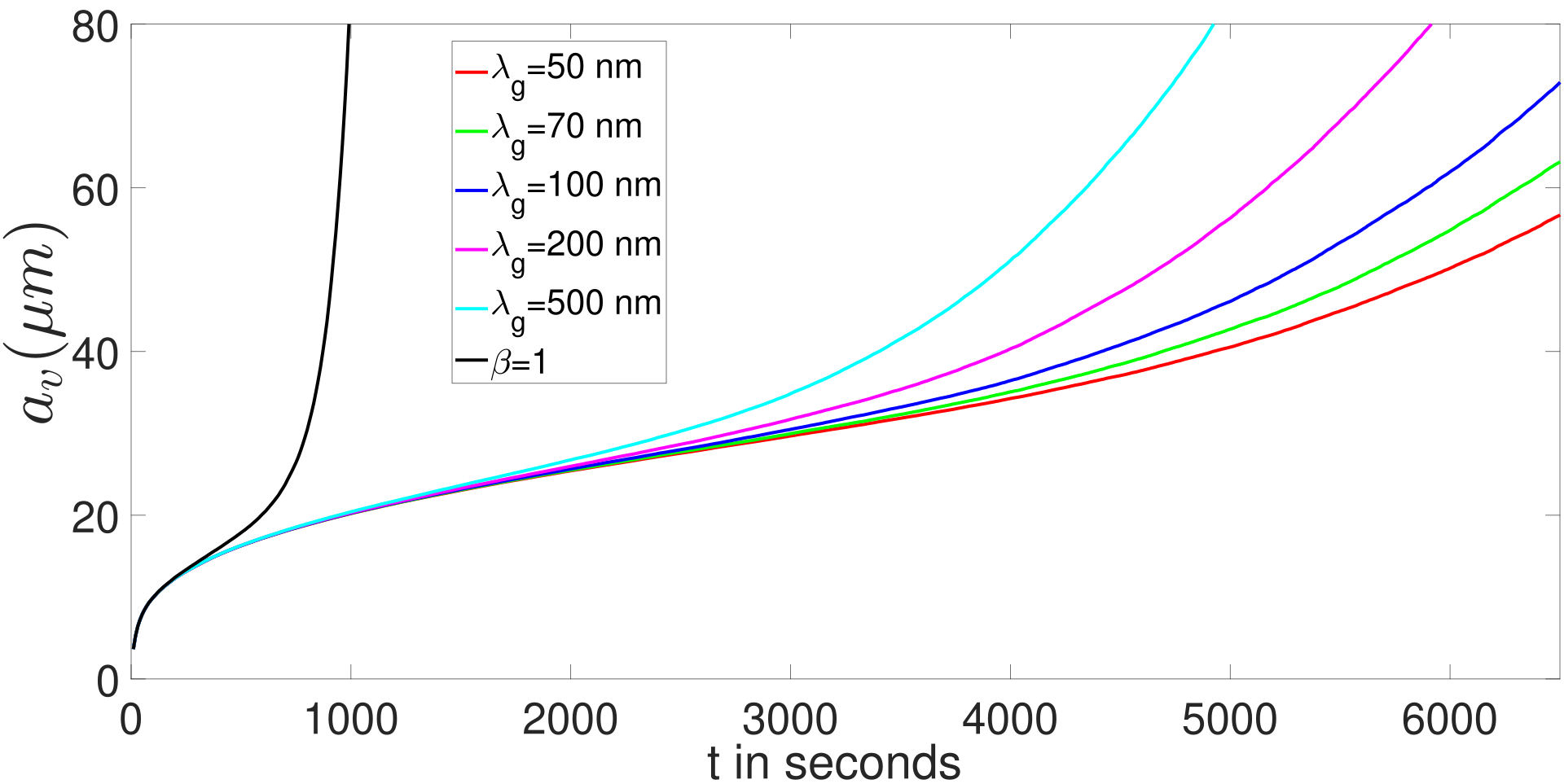}
\caption{The mean drop radius $a_v$ is shown for the full calculation performed with several fixed values of the mean-free path of the air.  The mean-free path values $\lambda_g=50,70,100,200,500$ nm are chosen along with an ideal-collision rate, $\beta=1$, calculation for reference.}
\label{fig:meanpath_var}
\end{figure}

Hydrodynamic interactions that include breakdown of continuum of the gaseous media upon close approach are very important for collision dynamics of water droplets in clouds \citep{sundararajakumar1996non} and we have shown, in figure \ref{fig:factors_combined}, their significant impact on the average drop size evolution. The lubrication force of this hydrodynamic interaction is strongly influenced by the mean-free path, which is $70$ nm at standard temperature and pressure and can vary due to changes in the ambient temperature and pressure. To obtain better insight into the role of $\lambda_g$ on the evolution of the droplet distribution we perform a series of calculations where $\lambda_g$ (and by extension $\nu$) remain constant at specified values during the rise of the air parcel. The resulting mean-radius evolution is shown in figure \ref{fig:meanpath_var} along with the reference case with the ideal collision rate, i.e.,  $\beta=1$. Although the results for mean-free paths ranging from 50-500 nm all exhibit much slower growth than obtained with the ideal collision rate, it is evident that the growth rate increases significantly with increasing $\lambda_g$.  Under typical terrestrial conditions  $\lambda_g$ is about 70-100 nm and $\beta$ is small. It can range from $0.1$ to $0.3$ for turbulent collisions and fall below $0.01$ for differential-sedimentation dominated collisions \citep{dhanasekaran2021collision_a}. 

In addition to informing the behaviour of terrestrial clouds at different altitudes, the simulation results with various $\lambda_g$ can provide insight into the droplet size evolution in extra-terrestrial atmospheres. For sulphuric acid precipitation on Venus and hydrocarbon rain on Titan, the Hamaker constants quantifying the van der Waals drop-drop attraction forces are about the same as for water droplets on Earth. The surface conditions on these planetary bodies indicate that the mean-free paths will be much smaller than that on earth\citep{williams2018dating}. However, significant precipitation activity can occur in Venus at altitudes of a few tens of kilometres\citep{gao2014bimodal}. \cite{perron2006valley} discuss the possibility of precipitation on Titan through the convective rise of air parcels up to heights of 40 kilometres. At these high altitudes the mean-free path can become large enough for non-continuum hydrodynamics to be significant. Hence, in figure \ref{fig:meanpath_var} we have performed calculations with $\lambda_g$ over a large range, including significantly larger values than those encountered in terrestrial clouds.

\begin{figure}
\hspace*{-0cm}
\includegraphics[scale=0.3]{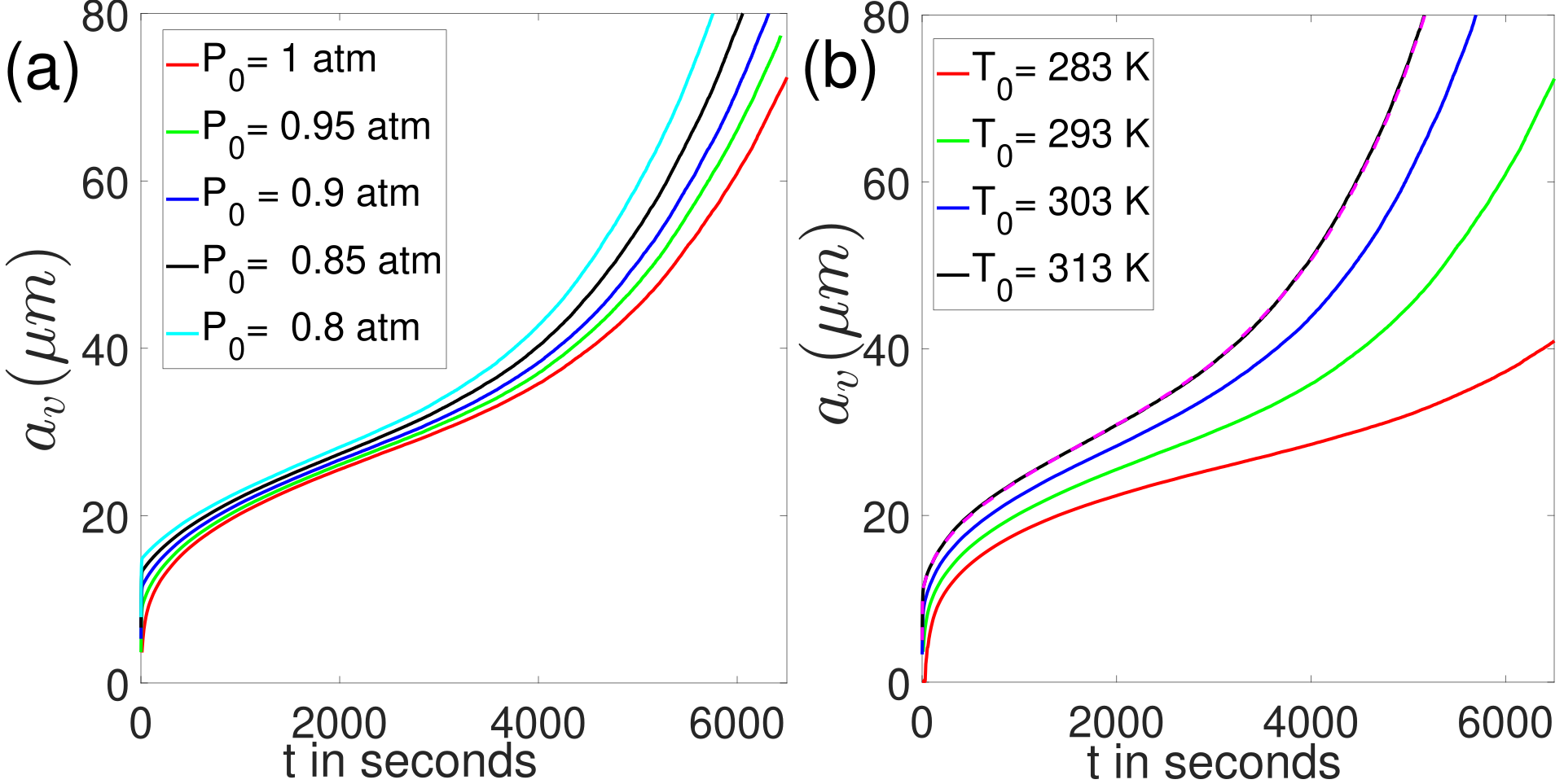}
\caption{Impact of varying initial conditions of air parcels interacting with the ground conditions on the time evolution of $a_v$. Figure (a) considers different initial pressures, from $P_0=1$ to 0.8 atmospheres. In figure (b), we vary the initial temperature from $T_0=283$ to 313 $K$. The dashed line corresponds to evolution with $T_0=313 K$ but $\lambda_g$ given the temporal dependence that would arise for a value of $T_0=293 K$.}
\label{fig:var_T_et_P}
\end{figure}

The range of mean-free path values $\lambda_g$ experienced by a rising cloud packet will depend on the ground pressure and temperature. Thus, as a further exploration of the effects of $\lambda_g$, we consider some typical values of $T_0$ and $P_0$ and show the resulting evolution of $a_v$ in figure \ref{fig:var_T_et_P}. Lower $P_0$ (figure \ref{fig:var_T_et_P} (a)) and higher $T_0$ (figure \ref{fig:var_T_et_P} (b)) lead to lower $\lambda_g$  values which, in turn, cause accelerated growth rate due decreased lubrication resistance to collision.

While changing $P_0$ predominantly alters $\lambda_g$ increases in $T_0$ also significantly enhance the water vapour carrying capacity of the air. Starting off with $100\%$ relative humidity the droplets from warmer ground can grow to larger sizes through condensation alone. If condensation can grow the droplets by an extra $10\%$, the ideal turbulent collision rate which is proportional to $(a_1+a_2)^3$, increases by $30\%$ and a similar enhancement occurs for the ideal differential sedimentation collision rate. To isolate the effect of changes in temperature on the water vapor content of the air, we perform a calculation with $T_0=313 K$ but values $\lambda_g$ based on a rising packet starting with $T_0=293 K$. The resulting evolution of $a_v$, which is shown as the dashed line in figure \ref{fig:var_T_et_P} (b), nearly tracks the simulation performed at $T_0=313 K$ indicating that temperature primarily influences drop growth through its effect on the water vapour carrying capacity of the air.  On the other hand the changes in droplet growth with pressure in figure \ref{fig:var_T_et_P} (a) are due to its effect on the mean-free path and the resulting non-continuum lubrication forces. 


\section{Conclusion}\label{sec:conclusion}

Previously unexplored microphysical influences on cloud droplet dynamics have been analysed in our study. We have considered a collision rate that properly couples turbulence with differential sedimentation and includes non-continuum hydrodynamics to analyse growth through the 'size-gap', where it is not fully understood how droplet growth occurs \citep{grabowski2013growth}. We also considered the role of droplet polydispersity resulting from  turbulence induced mixing of droplets and water vapour fluctuations. The relative importance of all of these mechanisms has been assessed.

Condensation was shown to have an important role in the early evolution washing away any memory of the initial distribution of micron-sized drops. The polydispersity induced by different histories of condensation, arising due to turbulent fluctuation of local water vapour content, was shown to enable crossing of the 'size-gap' in figure \ref{fig:fluc_Var_evol}. A large amount of condensation, due to enhanced water vapour content in the air parcel, was shown in figure \ref{fig:var_T_et_P} (b) to increase collisions by creating bigger drops in the 'size-gap'.

Turbulent collisions were shown in figure \ref{fig:eps_wv0_Var_evol} to be sufficient in crossing the 'size-gap' without the necessity for any variations in the supersaturation of the gas. The inclusion of polydispersity induced by variable condensation rates did not result in a simple additive contribution to growth. While the polydispersity due to water vapor fluctuations enhanced the growth rate at modest turbulence levels, a small retardation was observed in the turbulent dominated cases due to the nature of the hydrodynamic interactions and the weakening of inertial clustering.

The critical role of hydrodynamic interactions, that include the breakdown of continuum on close approach of droplets, has been demonstrated in figures \ref{fig:factors_combined} and \ref{fig:meanpath_var}. We explored the more moderate impact of inertial clustering through evolution of some important parameters in figures \ref{fig:gr_eps_wv01_Var_evol} and \ref{fig:gr_stsv_eps_wv01_Var_evol}. Crossing the 'size-gap' through turbulent collisions occurring in certain in high shear rate regions was shown to not be an effective route. Very few droplets grow to large sizes due to intermittency of the turbulent velocity gradient field. For those that do grow large, the differential sedimentation velocity induced is not large enough to make a significant impact on the evolution. This weaker than expected gravity induced velocity is due to the non-linear drag force and the significant reduction in the collision efficiency of large drops.

The results of this study can be extended to model industrial reactors and extra-terrestrial weather. For this purpose we have extensively spanned the $\lambda_g$ parameter space in figure \ref{fig:meanpath_var}. We have varied initial temperature and pressure in figure \ref{fig:var_T_et_P} to capture different ground conditions on Earth  as well providing insight into the interplay of the various mechanisms in shaping the droplet size evolution.

\acknowledgments
This work was supported by NSF grants 1435953 and 2206851.  The authors thank Raymond Shaw for useful discussions.


%
%
\datastatement
The simulation was performed using MATLAB software and the code along with the raw data have been uploaded to the Open Science Framework repository. The publicly accessible link is: https://osf.io/7h5sp/

%






%



 \bibliographystyle{ametsocV6}
 \bibliography{refer3_cloud}

\begin{thebibliography}{42}
\providecommand{\natexlab}[1]{#1}
\providecommand{\url}[1]{\texttt{#1}}
\renewcommand{\UrlFont}{\rmfamily}
\providecommand{\urlprefix}{URL }
\expandafter\ifx\csname urlstyle\endcsname\relax
  \providecommand{\doi}[1]{https://doi.org/\discretionary{}{}{}#1}\else
  \providecommand{\doi}{https://doi.org/\discretionary{}{}{}\begingroup \urlstyle{rm}\Url}\fi
\providecommand{\eprint}[2][]{\url{#2}}

\bibitem[{Ayala et~al.(2008)Ayala, Rosa, Wang,, and Grabowski}]{ayala2008effectsa}
Ayala, O., B.~Rosa, L.-P. Wang, and W.~Grabowski, 2008: Effects of turbulence on the geometric collision rate of sedimenting droplets. part 1. results from direct numerical simulation. \textit{New Journal of Physics}, \textbf{10~(7)}, 075\,015.

\bibitem[{Buesser and Pratsinis(2012)Buesser, and Pratsinis}]{buesser2012design}
Buesser, B., and S.~Pratsinis, 2012: Design of nanomaterial synthesis by aerosol processes. \textit{Annual review of chemical and biomolecular engineering}, \textbf{3}, 103--127.

\bibitem[{Chandrakar et~al.(2016)Chandrakar, Cantrell, Chang, Ciochetto, Niedermeier, Ovchinnikov, Shaw,, and Yang}]{chandrakar2016aerosol}
Chandrakar, K.~K., W.~Cantrell, K.~Chang, D.~Ciochetto, D.~Niedermeier, M.~Ovchinnikov, R.~A. Shaw, and F.~Yang, 2016: Aerosol indirect effect from turbulence-induced broadening of cloud-droplet size distributions. \textit{Proceedings of the National Academy of Sciences}, \textbf{113~(50)}, 14\,243--14\,248.

\bibitem[{Chandrakar et~al.(2024)Chandrakar, Morrison, Grabowski,, and Lawson}]{chandrakar2024turbulence}
Chandrakar, K.~K., H.~Morrison, W.~W. Grabowski, and R.~P. Lawson, 2024: Are turbulence effects on droplet collision--coalescence a key to understanding observed rain formation in clouds? \textit{Proceedings of the National Academy of Sciences}, \textbf{121~(27)}, e2319664\,121.

\bibitem[{Chaumat and Brenguier(2001)Chaumat, and Brenguier}]{chaumat2001droplet}
Chaumat, L., and J.-L. Brenguier, 2001: Droplet spectra broadening in cumulus clouds. part ii: Microscale droplet concentration heterogeneities. \textit{Journal of the atmospheric sciences}, \textbf{58~(6)}, 642--654.

\bibitem[{Chen et~al.(2018)Chen, Yau,, and Bartello}]{chen2018turbulence}
Chen, S., M.~Yau, and P.~Bartello, 2018: Turbulence effects of collision efficiency and broadening of droplet size distribution in cumulus clouds. \textit{Journal of the Atmospheric Sciences}, \textbf{75~(1)}, 203--217.

\bibitem[{Chun and Koch(2005)Chun, and Koch}]{chun2005coagulation}
Chun, J., and D.~Koch, 2005: Coagulation of monodisperse aerosol particles by isotropic turbulence. \textit{Physics of Fluids}, \textbf{17~(2)}, 027\,102.

\bibitem[{Chun et~al.(2005)Chun, Koch, Rani, A,, and Collins}]{chun2005clustering}
Chun, J., D.~L. Koch, S.~L. Rani, A.~A, and L.~R. Collins, 2005: Clustering of aerosol particles in isotropic turbulence. \textit{Journal of Fluid Mechanics}, \textbf{536}, 219--251.

\bibitem[{Clift et~al.(2005)Clift, Grace,, and Weber}]{clift2005bubbles}
Clift, R., J.~R. Grace, and M.~E. Weber, 2005: \textit{Bubbles, drops, and particles}. Courier Corporation.

\bibitem[{Davis(1984)}]{davis1984rate}
Davis, R., 1984: The rate of coagulation of a dilute polydisperse system of sedimenting spheres. \textit{Journal of Fluid Mechanics}, \textbf{145}, 179--199.

\bibitem[{Dhanasekaran and Koch(2022)Dhanasekaran, and Koch}]{dhanasekaran2022model}
Dhanasekaran, J., and D.~L. Koch, 2022: Model for the radial distribution function of polydisperse inertial spheres settling in homogeneous, isotropic turbulence. \textit{Physical Review Fluids}, \textbf{7~(10)}, 104\,602.

\bibitem[{Dhanasekaran et~al.(2021{\natexlab{a}})Dhanasekaran, Roy,, and Koch}]{dhanasekaran2021collision_a}
Dhanasekaran, J., A.~Roy, and D.~L. Koch, 2021{\natexlab{a}}: Collision rate of bidisperse, hydrodynamically interacting spheres settling in a turbulent flow. \textit{Journal of Fluid Mechanics}, \textbf{912}.

\bibitem[{Dhanasekaran et~al.(2021{\natexlab{b}})Dhanasekaran, Roy,, and Koch}]{dhanasekaran2021collision_b}
Dhanasekaran, J., A.~Roy, and D.~L. Koch, 2021{\natexlab{b}}: Collision rate of bidisperse spheres settling in a compressional non-continuum gas flow. \textit{Journal of Fluid Mechanics}, \textbf{910}.

\bibitem[{Dhariwal and Bragg(2018)Dhariwal, and Bragg}]{dhariwal2018small}
Dhariwal, R., and A.~Bragg, 2018: Small-scale dynamics of settling, bidisperse particles in turbulence. \textit{Journal of Fluid Mechanics}, \textbf{839}, 594--620.

\bibitem[{Falkovich and Pumir(2007)Falkovich, and Pumir}]{falkovich2007sling}
Falkovich, G., and A.~Pumir, 2007: Sling effect in collisions of water droplets in turbulent clouds. \textit{Journal of the Atmospheric Sciences}, \textbf{64~(12)}, 4497--4505.

\bibitem[{Feingold et~al.(1999)Feingold, Cotton, Kreidenweis,, and Davis}]{feingold1999impact}
Feingold, G., W.~R. Cotton, S.~M. Kreidenweis, and J.~T. Davis, 1999: The impact of giant cloud condensation nuclei on drizzle formation in stratocumulus: Implications for cloud radiative properties. \textit{Journal of the atmospheric sciences}, \textbf{56~(24)}, 4100--4117.

\bibitem[{Gao et~al.(2014)Gao, Zhang, Crisp, Bardeen,, and Yung}]{gao2014bimodal}
Gao, P., X.~Zhang, D.~Crisp, C.~G. Bardeen, and Y.~L. Yung, 2014: Bimodal distribution of sulfuric acid aerosols in the upper haze of venus. \textit{Icarus}, \textbf{231}, 83--98.

\bibitem[{Grabowski and Wang(2009)Grabowski, and Wang}]{grabowski2009diffusional}
Grabowski, W.~W., and L.-P. Wang, 2009: Diffusional and accretional growth of water drops in a rising adiabatic parcel: effects of the turbulent collision kernel. \textit{Atmospheric Chemistry and Physics}, \textbf{9~(7)}, 2335--2353.

\bibitem[{Grabowski and Wang(2013)Grabowski, and Wang}]{grabowski2013growth}
Grabowski, W.~W., and L.-P. Wang, 2013: Growth of cloud droplets in a turbulent environment. \textit{Annual review of fluid mechanics}, \textbf{45}, 293--324.

\bibitem[{Ireland et~al.(2016a)Ireland, Bragg,, and Collins}]{ireland2016aeffect}
Ireland, P., A.~Bragg, and L.~Collins, 2016a: The effect of reynolds number on inertial particle dynamics in isotropic turbulence. part 1. simulations without gravitational effects. \textit{Journal of Fluid Mechanics}, \textbf{796}, 617--658.

\bibitem[{Jennings(1988)}]{jennings1988mean}
Jennings, S., 1988: The mean free path in air. \textit{Journal of Aerosol Science}, \textbf{19~(2)}, 159--166.

\bibitem[{Koch and Pope(2002)Koch, and Pope}]{koch2002coagulation}
Koch, D.~L., and S.~B. Pope, 2002: Coagulation-induced particle-concentration fluctuations in homogeneous, isotropic turbulence. \textit{Physics of Fluids}, \textbf{14~(7)}, 2447--2455.

\bibitem[{Kostinski and Shaw(2005)Kostinski, and Shaw}]{kostinski2005fluctuations}
Kostinski, A.~B., and R.~A. Shaw, 2005: Fluctuations and luck in droplet growth by coalescence. \textit{Bulletin of the American Meteorological Society}, \textbf{86~(2)}, 235--244.

\bibitem[{Kulmala et~al.(1997)Kulmala, Rannik, Zapadinsky,, and Clement}]{kulmala1997effect}
Kulmala, M., {\"U}.~Rannik, E.~L. Zapadinsky, and C.~F. Clement, 1997: The effect of saturation fluctuations on droplet growth. \textit{Journal of aerosol science}, \textbf{28~(8)}, 1395--1409.

\bibitem[{Lasher-trapp et~al.(2005)Lasher-trapp, Cooper,, and Blyth}]{lasher2005broadening}
Lasher-trapp, S.~G., W.~A. Cooper, and A.~M. Blyth, 2005: Broadening of droplet size distributions from entrainment and mixing in a cumulus cloud. \textit{Quarterly Journal of the Royal Meteorological Society: A journal of the atmospheric sciences, applied meteorology and physical oceanography}, \textbf{131~(605)}, 195--220.

\bibitem[{Lasher-Trapp et~al.(2001)Lasher-Trapp, Knight,, and Straka}]{lasher2001early}
Lasher-Trapp, S.~G., C.~A. Knight, and J.~M. Straka, 2001: Early radar echoes from ultragiant aerosol in a cumulus congestus: Modeling and observations. \textit{Journal of the atmospheric sciences}, \textbf{58~(23)}, 3545--3562.

\bibitem[{Li et~al.(2017)Li, Brandenburg, Haugen,, and Svensson}]{li2017eulerian}
Li, X.-Y., A.~Brandenburg, N.~E. Haugen, and G.~Svensson, 2017: Eulerian and l agrangian approaches to multidimensional condensation and collection. \textit{Journal of Advances in Modeling Earth Systems}, \textbf{9~(2)}, 1116--1137.

\bibitem[{Li et~al.(2018)Li, Brandenburg, Svensson, Haugen, Mehlig,, and Rogachevskii}]{li2018effect}
Li, X.-Y., A.~Brandenburg, G.~Svensson, N.~E. Haugen, B.~Mehlig, and I.~Rogachevskii, 2018: Effect of turbulence on collisional growth of cloud droplets. \textit{Journal of the Atmospheric Sciences}, \textbf{75~(10)}, 3469--3487.

\bibitem[{Li et~al.(2019)Li, Svensson, Brandenburg,, and Haugen}]{li2019cloud}
Li, X.-Y., G.~Svensson, A.~Brandenburg, and N.~E. Haugen, 2019: Cloud-droplet growth due to supersaturation fluctuations in stratiform clouds. \textit{Atmospheric Chemistry and Physics}, \textbf{19~(1)}, 639--648.

\bibitem[{Peng et~al.(2002)Peng, Lohmann, Leaitch, Banic,, and Couture}]{peng2002cloud}
Peng, Y., U.~Lohmann, R.~Leaitch, C.~Banic, and M.~Couture, 2002: The cloud albedo-cloud droplet effective radius relationship for clean and polluted clouds from race and fire. ace. \textit{Journal of Geophysical Research: Atmospheres}, \textbf{107~(D11)}, AAC--1.

\bibitem[{Perron et~al.(2006)Perron, Lamb, Koven, Fung, Yager,, and {\'A}d{\'a}mkovics}]{perron2006valley}
Perron, J.~T., M.~P. Lamb, C.~D. Koven, I.~Y. Fung, E.~Yager, and M.~{\'A}d{\'a}mkovics, 2006: Valley formation and methane precipitation rates on titan. \textit{Journal of Geophysical Research: Planets}, \textbf{111~(E11)}.

\bibitem[{Saffman and Turner(1956)Saffman, and Turner}]{saffman1956collision}
Saffman, P., and J.~Turner, 1956: On the collision of drops in turbulent clouds. \textit{Journal of Fluid Mechanics}, \textbf{1~(1)}, 16--30.

\bibitem[{Siebesma et~al.(2003)}]{siebesma2003large}
Siebesma, A.~P., and Coauthors, 2003: A large eddy simulation intercomparison study of shallow cumulus convection. \textit{Journal of the Atmospheric Sciences}, \textbf{60~(10)}, 1201--1219.

\bibitem[{Smoluchowski(1918)}]{smoluchowski1918versuch}
Smoluchowski, M.~v., 1918: Versuch einer mathematischen theorie der koagulationskinetik kolloider l{\"o}sungen. \textit{Zeitschrift f{\"u}r physikalische Chemie}, \textbf{92~(1)}, 129--168.

\bibitem[{Sundararajakumar and Koch(1996)Sundararajakumar, and Koch}]{sundararajakumar1996non}
Sundararajakumar, R., and D.~L. Koch, 1996: Non-continuum lubrication flows between particles colliding in a gas. \textit{Journal of Fluid Mechanics}, \textbf{313}, 283--308.

\bibitem[{Sutherland(1893)}]{sutherland1893lii}
Sutherland, W., 1893: Lii. the viscosity of gases and molecular force. \textit{The London, Edinburgh, and Dublin Philosophical Magazine and Journal of Science}, \textbf{36~(223)}, 507--531.

\bibitem[{Van~Dongen and Ernst(1988)Van~Dongen, and Ernst}]{van1988scaling}
Van~Dongen, P., and M.~Ernst, 1988: Scaling solutions of smoluchowski's coagulation equation. \textit{Journal of Statistical Physics}, \textbf{50}, 295--329.

\bibitem[{Warner(1970)}]{warner1970microstructure}
Warner, J., 1970: The microstructure of cumulus cloud. part iii. the nature of the updraft. \textit{Journal of the Atmospheric Sciences}, \textbf{27~(4)}, 682--688.

\bibitem[{Westbrook et~al.(2004)Westbrook, Ball, Field,, and Heymsfield}]{westbrook2004theory}
Westbrook, C., R.~Ball, P.~Field, and A.~Heymsfield, 2004: Theory of growth by differential sedimentation, with application to snowflake formation. \textit{Physical Review E}, \textbf{70~(2)}, 021\,403.

\bibitem[{Wilkinson et~al.(2006)Wilkinson, Mehlig,, and Bezuglyy}]{wilkinson2006caustic}
Wilkinson, M., B.~Mehlig, and V.~Bezuglyy, 2006: Caustic activation of rain showers. \textit{Physical review letters}, \textbf{97~(4)}, 048\,501.

\bibitem[{Williams et~al.(2018)Williams, van~der Bogert, Pathare, Michael, Kirchoff,, and Hiesinger}]{williams2018dating}
Williams, J.-P., C.~H. van~der Bogert, A.~V. Pathare, G.~G. Michael, M.~R. Kirchoff, and H.~Hiesinger, 2018: Dating very young planetary surfaces from crater statistics: A review of issues and challenges. \textit{Meteoritics \& Planetary Science}, \textbf{53~(4)}, 554--582.

\bibitem[{Xue et~al.(2008)Xue, Wang,, and Grabowski}]{xue2008growth}
Xue, Y., L.-P. Wang, and W.~W. Grabowski, 2008: Growth of cloud droplets by turbulent collision--coalescence. \textit{Journal of the Atmospheric Sciences}, \textbf{65~(2)}, 331--356.

\end{thebibliography}

\end{document}